# Study of Auto-igniting Spray Flame in Vitiated Swirling Hot Coflow using flamelet generated model

Zafar Alam[1], Bharat Bhatia[1,2] and Ashoke De[1,2,a]

[1]*Department of Aerospace Engineering, Indian Institute of Technology Kanpur, 208016, Kanpur, India.*

[2]*Current Affiliation: Department of Mechanical Engineering, Eindhoven University of Technology, Eindhoven-5600MB, Netherlands*

[3]*Department of Sustainable Energy Engineering, Indian Institute of Technology Kanpur, 208016, Kanpur, India.*

Swirl-stabilized auto-igniting spray flames are essential for designing efficient and clean combustion systems. The present study performs large eddy simulations (LES) of the dilute auto-igniting methanol flame in a vitiated, hot coflow of varying swirl intensities. The six-dimensional Flamelet Generated Manifold (FGM) technique is used to solve the reactive flow accurately and economically. The swirl numbers ($S_N$), i.e. 0.2, 0.6, 1.0, and 1.4, are used to assess their effect on auto-ignition and flame stability. At lower to moderate swirl numbers ($S_N$ =0.2, 0.6), the increase in swirl is found to increase the lift-off height. Beyond the critical swirl number ($S_N$=0.6), the lift-off height drops. Also, the time-averaged flame structure transitions from a tubular-like flame into a uniformly distributed combustion region at these high swirl numbers. It also results in a more compact flame for the higher swirl numbers. These effects on flame dynamics are analyzed in detail using the mean gas-phase flow field distribution, particle statistics, and proper orthogonal decomposition (POD).

## I. INTRODUCTION

Swirl combustion is a significant factor in the design of combustion systems. There has been a lot of research on the effect of swirl on combustion dynamics focused on the performance and efficiency of gas turbine engines[1]. A swirl adds the tangential momentum to the gaseous fuel/oxidizer entering the combustion chamber, promoting fuel-oxidizer mixing and, hence, efficient combustion. The parameters governing the swirl depend on whether it is applied to fuel, oxidizer, or both. It also depends on the fuel type, the burner or combustor design, and the desired emissions limit. The increased residence time of the fuel/oxidizer mixture brought about by swirl addition helps to complete slow oxidation reactions. Also, introducing the right intensity of the swirl leads to forming vortical structures that stabilize the flame. This becomes an essential factor in the design of gas-turbine powerplants or aircraft engines, where steady performance is critical. Past attempts have delved into these design aspects, focusing on mixing, stability, efficiency, and complete combustion[2-4].

______________________________

a) Electronic mail: ashoke@iitk.ac.in.

Generally, the swirl strength is measured in terms of swirl numbers ($S_N$). It is found that the vortical structure formed in the primary zone introduces a central recirculation zone (CRZ) at higher $S_N$. These zones recirculate the hot exhaust gas back upstream, enhancing better combustion at comparatively lower temperatures while stabilizing the flame. It has been reported that increasing the swirl strength reduces the carbon monoxide (CO) and nitrogen oxides ($NO_x$) formation[5]. With the swirl flames comes the challenge of swirl injector design as well[6-8]. On the contrary to high swirl, low swirl injectors are preferable for fuels involving hydrogen ($H_2$) because of their more tolerance to flashbacks[8]. Other studies have also where the premixed charge is injected into the chamber[12,14]. However, in the case of Lean-Premixed Combustion (LPM) type combustors where a premixed gaseous mixture is fed into the combustor, the combustion instability has been a common problem hindering its development[11]. There have been efforts to control the instability through active modulation. The thermoacoustic pressure can be suppressed using a closed-loop active control system that modulates the airflow and, thus, the mixing and combustion [9,10,13]. This method is found to stabilize the flame while also reducing the NOx emissions [9].

When the fuel involved in combustion is in the form of liquid droplets, either it is pre-vaporized before introducing into the combustion chamber (referred to as lean-premixed-prevaporized (LPP) technology)[15] or injected directly in the chamber (referred to as lean-direct injection (LDI) technology)[16]. These methods have been vastly researched and are currently used in aircraft engines. Compared to LPM or LPP, thermo-acoustic instability is less dominant in LDI. Instead, autoignition is an essential phenomenon in LDI. The LDI-based combustor designs may also use a multi-point injection system for better mixing[17]. These types of combustor systems generally involve fuel-air combustion in a swirl condition.

Many computational studies have simulated the spray combustion in combustors considering swirl flows. Jones et al. [18] used the joint probability density function to solve the scalars in the LES simulation of swirl-stabilized kerosene spray flame. Dinesh et al. [19] used the Smagorinsky eddy viscosity model for momentum and scalar equations in the LES simulation of a coannular swirl jet. They witnessed CRZ and precessing vortex cores (PVC) in all their cases. In a similar computational study, Sankaran & Menon[12] captured the recirculation bubble due to vortex breakdown and showed the impact of swirl improves fuel-air mixing by increasing droplet dispersion and hastening the disintegration of vortices in gas turbine combustors scenario.

However, canonical burners have been developed to understand the underlying combustion phenomena better and remove the extra complexity of combustor design [20]. The study of auto-igniting flames, such as the methanol spray flame in hot coflow [20], has gained interest concerning autoignition in combustors. There have been computational studies on this burner primarily focused on model development and validation of the flame and/or spray characteristics [21-24].

An essential aspect of the autoigniting flame is its liftoff height, which depends on the vortical structures and, hence, the mixing. But, it will be interesting to know the response of an autoigniting flame to the swirl flow. Swirling in a coflow can result in the PVCs affecting the flame ignition and liftoff height. To understand this, we have conducted a computational LES study on the dilute spray burner in a swirling hot coflow of varying intensities and performed proper orthogonal decomposition (POD) on the flowfield to investigate the vortical and flame structures. We use the extended-FGM for combustion modeling of this type of burner as validated in the present authors' previous study [24]. The details of the solver are provided in Section II. The different cases with varying swirl intensities are explained in Section III, followed by an analysis of the mean flow results, particle statistics, and POD modes obtained in Section IV, and finally, the conclusion in Section V.

## II. NUMERICAL METHODOLOGY

The simulation process involves tracking two distinct phases: the liquid and gas phases. The liquid phase, made up of tiny droplets, is tracked using a Lagrangian approach. On the other hand, the gas phase, which includes the air and vaporized fuel, is modeled using an Eulerian approach, treating the gas as a continuous fluid. The chemical reactions between the vaporized fuel and the air are described using a simplified Flamelet Generated Manifold (FGM) model. The mathematical equations governing these two phases and the FGM model will be explained in detail in the following sections.

### A. Basic Governing Equations

The continuity equation, momentum equation, energy equation, and scalar equations related to the FGM model are among the filtered transport equations that are applied to the Eulerian (gas) phase. Filtering operations for unweighted and density-weighted ensemble averages are represented by the overbar and tilde ($\sim$), respectively. Source terms that characterize the transfer between the two phases are included in the continuity and momentum equations, which are given in [24]

$$\frac{\partial \bar{\rho}}{\partial t} + \frac{\partial}{\partial x_i}(\bar{\rho} u_i) = \overline{S_{mass}} \tag{1}$$

$$\frac{\partial \bar{\rho} u_i}{\partial t} + \frac{\partial}{\partial x_j}(\bar{\rho} u_i u_j) = -\frac{\partial \bar{p}}{\partial x_i} + \bar{\rho} g_i + \frac{\partial}{\partial x_j}(\bar{\tau}_{ij}) + \frac{\partial}{\partial x_j}(\overline{\rho u_i u_j} - \bar{\rho} u_i u_j) + \overline{S_{mom,i}} \tag{2}$$

In Eq. 2, the term $\bar{\tau}_{ij}$ represents the stress tensor, encompassing both laminar and subgrid-scale effects.

$$\bar{\tau}_{ij} = 2\bar{\rho} \nu_{eff}\left(S_{ij} - \frac{1}{3}\delta_{ij} S_{ii}\right) \tag{3}$$

$$S_{ij} = \frac{1}{2}\left(\frac{\partial u_i}{\partial x_j} + \frac{\partial u_j}{\partial x_i}\right) \tag{4}$$

Where $v_{eff}$ is the effective kinematic viscosity. In this study, the dynamic K-equation model, a one-equation eddy viscosity SGS model, is applied. The turbulent kinetic energy is calculated using the following equation[25-27]:

$$\frac{\partial(\overline{\rho}k_{sgs})}{\partial t} + \frac{\partial(\overline{\rho}u_i k_{sgs})}{\partial x_i} = \frac{\partial}{\partial x_i}\left(\overline{\rho}(D + D_{sgs})\frac{\partial k_{sgs}}{\partial x_i}\right) - C_\varepsilon \frac{\overline{\rho}k_{sgs}^{3/2}}{\Delta} - \overline{\rho}\left(\tau_{ij}^{sgs} : \varepsilon\right) \quad (5)$$

$$D_{sgs} = C_k \sqrt{k_{sgs}} \Delta \quad (6)$$

$$\varepsilon : \tau_{ij}^{sgs} + C_\varepsilon \frac{k_{sgs}^{3/2}}{\Delta} = 0 \quad (7)$$

Where $D_{sgs}$ sub-grid scale diffusivity, $k_{sgs}$ is the sub-grid scale kinetic energy, and the double dot product is denoted by (:). $C_k$ and $C_\varepsilon$ constants are calculated based on dynamic formulation[25-27]. The governing partial differential equations are transformed into algebraic equations using a Finite Volume Method (FVM) implemented in the OpenFOAM-v1912[28]. The convective terms in the equations are discretized using a second-order Total Variation Diminishing (TVD) scheme, while the viscous terms are discretized using a second-order central scheme. The time-dependent terms are handled using the second-order Crank-Nicolson scheme with sufficiently small-time steps to ensure numerical stability and minimize numerical diffusion.

**B. Lagrangian Framework**

The near-spherical droplets produced in the atomization process are converted into Lagrangian particles, which act as point sources of energy, momentum, and mass for the previously described equations. The simulation involves many computational particles, each representing a group of physical droplets with identical properties. The text will refer to these particles interchangeably as particles or droplets. The equations governing the behavior of these particles incorporate sub-models for heat transfer, atomization, dispersion, and droplet collisions. The position and velocity of the particles are determined using the Basset-Boussinesq-Oseen (BBO) equation, which is given by:

$$\frac{dx_p}{dt} = u_p \quad (8)$$

$$m_p \frac{du_p}{dt} = F_D + F_G + F_T \quad (9)$$

In this equation, $u_p$, $m_p$, $x_p$ represent the velocity, mass, and position of each particle, respectively. The terms $F_G$ and $F_D$ represent the gravitational and drag forces acting on the particles, respectively; however, $F_T$ represents the turbulence effect modeled using the gradient dispersion model [29].

Lagrangian parcels exhibit physical phenomena such as evaporation, heat transfer, and droplet breakup. The heat transfer process is modeled using the Ranz-Marshall model developed by Ranz and Marshall[30]. The evaporation process is calculated using the Frossling correlation, which assumes the Lagrangian droplets are spherical[31]. Further details about the equations discussed in this section are provided in the appendix.

## C. Flamelet Generated Manifold (FGM)

The chemical reactions occurring after the injection of the liquid jet are represented as one-dimensional laminar flames, referred to as flamelets [33]. Assuming the flamelet assumptions apply to the resulting turbulent flame, only a limited governing variable must be solved for flame dynamics. The CHEM1D code[32] is employed to compute the flamelets under ambient pressure conditions. In order to fully characterize the one-dimensional flame structure to incorporate turbulence chemistry interaction closure, adaptive grid refinement is used to solve one-dimensional temperature, species, and flow equations with an equation of state (EOS). The ideal gas EOS has been applied to the current atmospheric flames. As the scalar dissipation value (or strain rate) gets closer to zero, the states that are closest to the chemical equilibrium state are achieved. The highest temperature of these flamelets is close to the adiabatic mean temperature. However, no reaction occurs when the mixing solution is produced. The flamelet properties, including species concentrations and chemical reactions, are determined as temperature and mixture fraction functions for methanol fuel and vitiated hot coflow oxidizer. These flamelets are calculated in a counter-flow configuration for 32 species, with 167 reactions in the chemical mechanism[33]. The flamelet calculations cover a range of conditions from near-equilibrium states at low strain rates to unsteady flames and, ultimately, pure mixing without reactions at high strain rates. Steady flamelets without secondary oxidizer range from near zero to around 3000s$^{-1}$ of strain rate; however, the steady flamelets with secondary oxidizer range from near zero to around 400s$^{-1}$ of strain rate[24]. The flamelet data, calculated initially in physical space, is transformed into a control-variable space for integration into the FGM table. For detailed equations, please refer to the appendix.

### *C1. Mixture Fraction ($Z_1$)*

Bilger et al.[34] defined one of the control variables as mixture fraction, given by,

$$Z_1 = \frac{Y_e - Y_e^{Ox}}{Y_e^F - Y_e^{Ox}}, \tag{10}$$

$Y_e$ and $M_w$ are the coupling function and molecular weight of carbon, hydrogen, and oxygen atoms.

### C2. Progress Variable ($Y_c$)

Another crucial control variable is the unscaled progress variable, denoted as $Y_c$, which quantifies the extent of combustion. This variable is a weighted sum of specific species, including $CO_2$, $H_2O$, and $H_2$, indicating combustion progress. The progress variable ($Y_c$) is calculated as[35-36].

$$Y_c = \frac{Y_{CO_2}}{M_{CO_2}} + \frac{Y_{H_2O}}{M_{H_2O}} + \frac{Y_{H_2}}{M_{H_2}}. \tag{11}$$

M represents the molar mass used as the weighting factor. The unscaled progress variable is normalized based on its minimum and maximum values. When considering four control variables, the progress variable's final value is influenced by enthalpy loss, mixture fraction, and a second mixture fraction that characterizes the mixing of the two oxidizer streams. For a detailed explanation, see Appendix (FGM Framework).

### C3. Enthalpy Deficit

The loss of normalized enthalpy is considered as follows[37]:

$$\eta = \frac{h - h_{ad}}{(1 - Z_1)(h_{Ox,\eta=1} - h_{Ox,\eta=0})} \tag{12}$$

The oxidizer enthalpy under adiabatic conditions is denoted by $h_{Ox,\eta=1}$, while the enthalpy corresponding to the maximum energy loss, determined by a specified minimum temperature, is represented by $h_{Ox,\eta=0}$. The minimum temperature for the hot coflow configuration is set to 1030K, whereas for the air jet, it is set to 268K.

### C4. Oxidizer Mixture fraction ($Z_2$)

Introducing a mixture fraction for the oxidizer allows the characterization of the mass fraction of the two oxidizer components within the mixture. In this study, the oxidizers are air and hot coflow. The oxidizer mixture fraction is defined according to the equation presented as follows[37]:

$$Z_2 = \frac{Y_{O_2} - Y_{O_2,HCF}}{Y_{O_2,Air} - Y_{O_2,HCF}} \tag{13}$$

Here, subscript HCF refers to hot coflow. Each flamelet is associated with a constant value of $Z_2$ at its oxidizer boundary, representing a non-reacting scalar. This value indicates the local mass fraction contributed by each oxidizer stream and the primary mixture fraction, $Z_1$.

### C5. Mixture Fraction and Progress Variable Variance

The scaled variance of the mixture fraction, ranging from 0 to 1, and the progress variables are calculated according to the following equations[37]:

$$\zeta_{PV} = \frac{\widetilde{Y_c''^2}}{\widetilde{Y_c}(1-\widetilde{Y_c})} \qquad \zeta_{PV} = \frac{\widetilde{Y_c''^2}}{\widetilde{Y_c}(1-\widetilde{Y_c})} \tag{14}$$

## D. Proper Orthogonal Decomposition (POD)

Proper orthogonal decomposition (POD) is a mathematical method used to identify and extract the most important patterns of variation within a dataset. It enables the breakdown of complex turbulent flows into simplified flow patterns based on chosen flow variables. These flow patterns are represented by orthogonal eigenvectors linked to corresponding eigenvalues, which can be used to reconstruct the entire flow field. A reduced-dimensional representation of the system can be achieved by focusing on a limited number of modes that capture most of the flow's variability. This technique is widely used to understand the fundamental mechanisms driving transport processes within the original flow[38-39].

The specific research focus, flow conditions, and intended application determine the selection of variables used for POD analysis. For example, studying compressible or reactive flows requires considering thermodynamic properties like temperature. In the context of autoigniting flames, the species OH, which is essential for ignition, can be employed to examine the formation of ignition kernels.

Initially, a variable corresponding to the relevant N x N covariance matrix A and the matrix of field values (encompassing all grid points) across a set of N snapshots is calculated.

$$A = X^T X \tag{15}$$

Then, the eigenvalue problem **A** is resolved, yielding eigenvalues $\lambda_k$ and eigenfunctions $\varphi_k$ $1 \leq k \leq N$:

$$A\varphi_k = \lambda_k \varphi_k \tag{16}$$

The energy associated with each mode is represented by the corresponding eigenfunction, ordered from highest to lowest eigenvalue. The projections of the data matrix **X** onto the eigenvectors yield the POD modes.

$$\phi_k = \varphi_k X \qquad 1 \leq k \leq N \tag{17}$$

$$X_{reconst} = \sum_{k=1}^{P} b_k \phi_k \tag{18}$$

The POD time coefficients, represented by $b_k$, change over time. They are obtained by projecting the time-varying flow fields onto the fixed POD modes. Essentially, these coefficients capture the temporal evolution of the flow field.

$$b_k = \Psi^T x_k \tag{19}$$

Now, POD modes $[\phi_1, \phi_2, \phi_3, \ldots]$ form a matrix $\Psi$ and, for a given $k^{th}$ snapshot, **x** represent the changing field of the covariance matrix. This study employs this methodology to investigate the formation of ignition kernels and flame expansion. The trace of the covariance matrix quantifies the average energy of the fluctuating LES velocity field. If u represents the velocity,

$$tr(A) = \langle (u'_i, u'_i) \rangle = \sum_{k=1}^{N} \lambda_k \tag{20}$$

Here, $\langle \cdot \rangle$ shows the operator of time-averaging. Moreover, by forming a covariance matrix, the velocity-temperature field ($u'_i$, $T'$) analysis is performed as:

$$tr(A) = \sum_{k=1}^{N} \lambda_k = \langle (u'_i, u'_i) \rangle + \gamma^2 \langle (T', T') \rangle \tag{21}$$

The $\gamma^2 = \langle (u'_i, u'_i) \rangle / \langle (T', T') \rangle$ introduced coefficient makes the two fluctuating fields consistent[40-41].

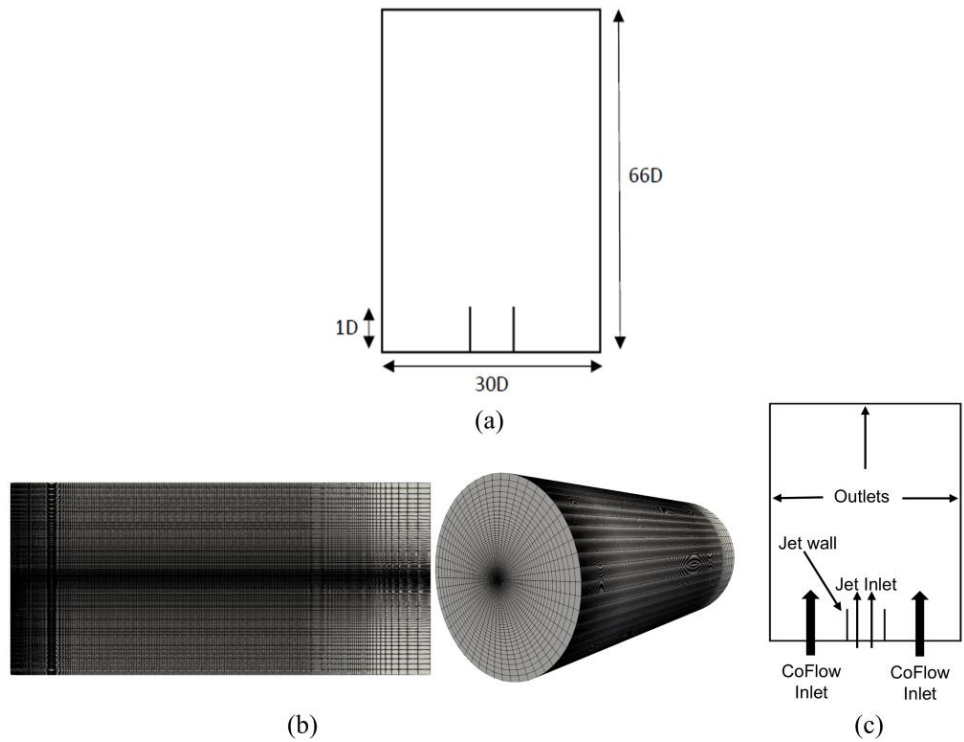

FIG. 1. (a) Schematics of the computational domain involved in the present study (b) Mesh used in the present study (c) Schematics of boundary conditions involved in the current study.

## III. NUMERICAL DETAILS

### A. Geometry

A detailed configuration and setup of the burner is available in the publication by O'Loughlin & Masri [42], which covers the experimental work. The design features a central carrier jet of cold air with a diameter of 4.6 mm, encircled by a hot outer coflow with a diameter of 197 mm, which issues combustion products from a hydrogen/air flame.

The computational domain, measuring 30D × 66D where D is the diameter of the center jet, corresponds to the burner used in the experiment and is depicted in Figure 1(a). An extension of 1D is applied for the jet entry. The center jet, fueled with methanol, is supplied at a velocity of 75 m/s, while the coflow has a temperature of 1430 K and a velocity of 3.5 m/s.

### B. Mesh

Figure 1(b) shows the mesh used in the present study, which consists of 2 million grid points, as determined from a previous study[24]. The cells near the nozzle have dimensions of approximately 0.110 mm × 0.115 mm × 0.240 mm along the axial, radial, and azimuthal axes, respectively. This mesh configuration was selected to ensure sufficient resolution for accurately capturing the flow and combustion phenomena in the simulation.

## C. Boundary Conditions

Figure 1(c) presents a schematic representation of the boundary conditions imposed on the domain. The diagram clearly illustrates how various boundary conditions are applied to different domain sections. The dilute methanol spray passing through the burner is subjected to the conditions mentioned in Table I. The coflow consists of the combustion products, primarily nitrogen, oxygen, and water vapor, along with trace amounts of other radicals, as shown in Table II.

TABLE I. shows the subjected conditions of dilute methanol spray as it passes through the burner.

| | |
|---|---|
| Coflow Temperature, $T_{coflow}$ | 1430K |
| Central Jet Temperature, $T_{jet}$ | 288K |
| Fuel loading | 0.295 |
| Coflow mixture fraction, $Z_{coflow}$ | 0 |
| Jet mixture fraction, $Z_{jet}$ | 0.080 |
| Coflow progress variable, $Y_{c,coflow}$ | 5.75 |
| Jet progress variable, $Y_{c,jet}$ | 0 |
| Coflow oxidizer mixture fraction, $Z_{2,coflow}$ | 0 |
| Jet oxidizer mixture fraction, $Z_{2,jet}$ | 1 |

TABLE II. Mole fraction of major species in the coflow at the given temperature.

| Temperature (K) | Equivalence ratio | N2 | O2 | H2O | H2 | O | OH |
|---|---|---|---|---|---|---|---|
| 1430K | 0.4 | 0.729 | 0.116 | 0.155 | $3.1 \times 10^{-7}$ | $4.9 \times 10^{-7}$ | $4.5 \times 10^{-5}$ |

TABLE III. Swirl Cases involved in the present study.

| Cases | Swirl Number ($S_N$) |
|---|---|
| Case-1 | 0.2 |
| Case-2 | 0.6 |
| Case-3 | 1.0 |
| Case-4 | 1.4 |

Table III shows the different swirl cases involved in the current study. For an annular swirler with a fixed vane angle θ, the swirl number ($S_N$), formulated by Syred & Beer[43], is defined by the following equation. This parameter quantifies the ratio of angular momentum to axial momentum in the flow, which is crucial in characterizing the degree of swirl imparted to the flow by the swirler.

$$S_N = \frac{2}{3} \left( \frac{1-\left(\frac{D_{hub}}{D_{sw}}\right)^3}{1-\left(\frac{D_{hub}}{D_{sw}}\right)^2} \right) \tan\theta \tag{22}$$

Here, $D_{sw}$ is the swirler diameter, and $D_{hub}$ is the swirler hub diameter.

By splitting its radius into ten intervals and the azimuthal direction into sixteen intervals, the spray inlet is separated into multiple injection patches with evenly spaced center points in the inlet jet-exit plane. The locations of experimental measurements serve as the primary motive for the radial divisions. While not too fine to permit the injection of parcels from every patch at every time step, the azimuthal distribution is homogeneous and sensitive enough to depict the circumferential uniformity of droplet injection while conforming to the experimental data of volume flux for differently sized droplets.

Each patch contains five overlapping sub-patches to measure the input mass flow of droplets categorized into various size bins. Based on previous experimental results, these size bins range from 0 to 10 µm, 10 to 20 µm, 20 to 30 µm, 30 to 40 µm, and 40 to 50 µm. The mass flow, mean velocities, and fluctuating root mean square (r.m.s) velocities are provided for each injector patch. The injector releases approximately 2 million parcels per second at an initial temperature of 288K, with the five overlapping sub-patches in each patch used to measure the mass flow across different droplet size bins.

## IV. RESULTS AND DISCUSSION
### A. GRID INDEPENDENCE AND VALIDATION

The quality and dependability of the simulation findings were determined by Bhatia et al.'s thorough validation and grid independence analysis in their study[24]. They comprehensively compared two distinct grid resolutions to verify that the mesh was sufficiently fine-tuned to capture the flow field's key characteristics. Although coarser than the finer alternative, the chosen mesh underwent careful refinement to ensure its ability to resolve fundamental flow dynamics with high precision. A helpful key tool for assessing grid resolution in simulations is the LES index of resolution quality. In accordance with Pope's criteria for a satisfactory mesh resolution, Celik et al.[44] suggested an index based on eddy viscosity, with a suggested value of 75%–80% or above. Bhatia et al.[24] examined the LES index field for coarse and fine grids and found that, regardless of grid size, values higher than 85% were consistently prevalent across the area. This demonstrates how reliable the index is for evaluating resolution quality across various grid configurations. The mesh resolution was specifically improved to roughly 0.110 mm in the axial direction, 0.115 mm in the radial direction, and 0.240 mm in the azimuthal direction close to the jet exit. These dimensions were carefully chosen to ensure proper resolution of the velocity and temperature gradients in the lift-off region of the flame.

The Mt2C flame was used as a reference case in a thorough grid sensitivity investigation by Bhatia et al.[24]. The study compared data from the two grid sizes to investigate how mesh refinement affects simulation accuracy. Lift-off height, velocity profiles, and temperature distribution were among the primary flow characteristics that were almost the same in both grids. This suggests that the coarser grid could accurately depict the fundamental physics of the flame. The validity of the coarser mesh was further supported by the remaining agreement between the radial profiles of axial velocity and temperature at two downstream locations and experimental data, as shown in their paper.

The refinement near the jet exit and throughout the domain ensured the grid was fine enough to capture the critical mixing and reaction zones, which are crucial for accurately simulating the flame's behavior. Bhatia et al.[24]'s results demonstrate that the mesh balances computational efficiency and solution accuracy. Given the negligible differences between the coarse and fine mesh results and the strong agreement with experimental data, the same coarser mesh has been selected for the simulations in this study. Their findings validate the chosen mesh and confirm that the flow physics and combustion characteristics have been captured with sufficient resolution. Therefore, mesh selection for detailed simulations is well-supported, ensuring that the computational model is efficient and reliable for capturing this study's intricate flow and flame dynamics.

**B. MEAN TEMPERATURE AND VELOCITY VARIATIONS**

The temperature and velocity profiles for four cases—S1, S2, S3, and S4-flames—with swirl numbers ($S_N$) of 0.2, 0.6, 1.0, and 1.4 are displayed at various axial points. Understanding how swirl affects an auto-igniting flame's dynamics is made more accessible by these profiles, which offer insightful information about the structure of jets and flames. Comparing swirl flames (S1-4) against those without swirl ($S_N = 0$, referred to as NS-flame from here onwards) allows a thorough examination of the swirl effect on flame dynamics.

As the flow moves downstream, the velocity profile in Figure 2 decreases and eventually flattens. The flow field velocity decreases when a swirl is added into the coflow. The swirling motion's enhanced turbulence is the main reason for higher energy dissipation in the shear layer region. Thus, the swirling flow further decreases the axial gas velocity as the tangential velocity component increases over the axial component with an increased swirl number.

Figure 3 shows the temperature profiles at downstream positions ranging from $x/D = 5$ to $x/D = 40$ for four methanol flame cases, where D represents the diameter of the central carrier jet. In all the scenarios, the swirl number progressively increases in Figure 2 while the hot coflow temperature remains constant at 1430K. This comparative analysis facilitates a detailed investigation into the effect of swirl on the temperature distribution along the flame axis, providing a clearer understanding of the role of coflow swirl on flame behavior.

The radial temperature profile with a significant dip for $x/D=5$ corresponds to the cold central carrier jet due to the evaporation of the liquid fuel droplets. At this stage, the effect of various swirl numbers on the temperature profile is not yet noticeable. By $x/D = 10$, the temperature shows a small increment beyond the surrounding hot coflow temperature, indicating some heat release. The coflow swirl, through the generation of turbulent eddies, promotes mixing and enhances the heat transfer between the central colder jet and the surrounding hot gases. It increases the central jet temperature for high swirl cases, as visible at $x/D=10$ for $S_N = 1.0$ and 1.4.

At x/D = 20, the core jet of all swirl flames exhibits a notable temperature increase, with peak values in the shear layers, suggesting efficient ignition in these areas. Because of the increased mixing, the temperature profile is more uniform in swirl flames than in NS flames. All of the flames show temperatures higher than the coflow temperature at downstream points of x/D = 30 and x/D = 40, suggesting the occurrence of the reaction. It can be noticed that the temperature profile falls when the swirl number is raised more than $S_N = 0.2$. It should be mentioned that the flames NS and S1 continue to have a developed potential jet-core region until x/D = 40, whereas the jet shear layer merges earlier (between x/D 30 to x/D 40) for the flames S2, S3, and S4. This is the reason for the drop in the middle of the temperature profile of NS and S1 flames. Added swirl to the NS flame increases the shear and the turbulence in the S1 flame, facilitating the mixing that raises the temperature profile marginally greater than that of the NS flame. On the other hand, the temperature profiles for flames S2 and S3 decrease and nearly match that of the NS flame. The temperature profile of the S4 flame with the maximum swirl is the lowest of all the flames. As the potential core region has ended before x/D = 30, these high swirl flames have a flatter temperature profile, and the flame covers a broader region.

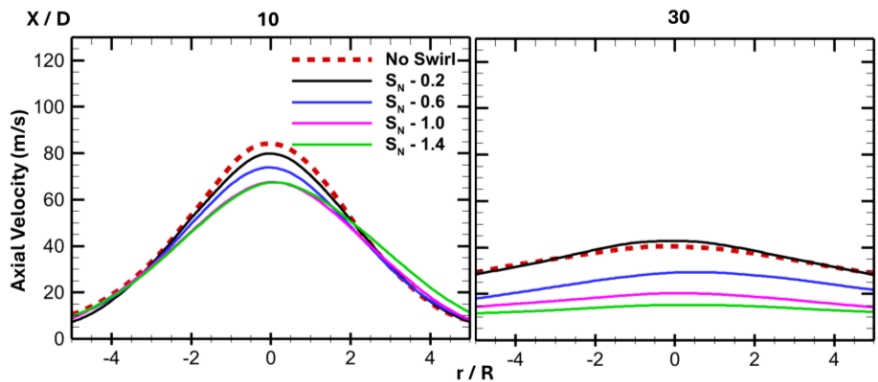

FIG 2. Radial distribution of axial velocity at x/D 10 and x/D 30 downstream locations for all swirl flames, compared to the no-swirl flame

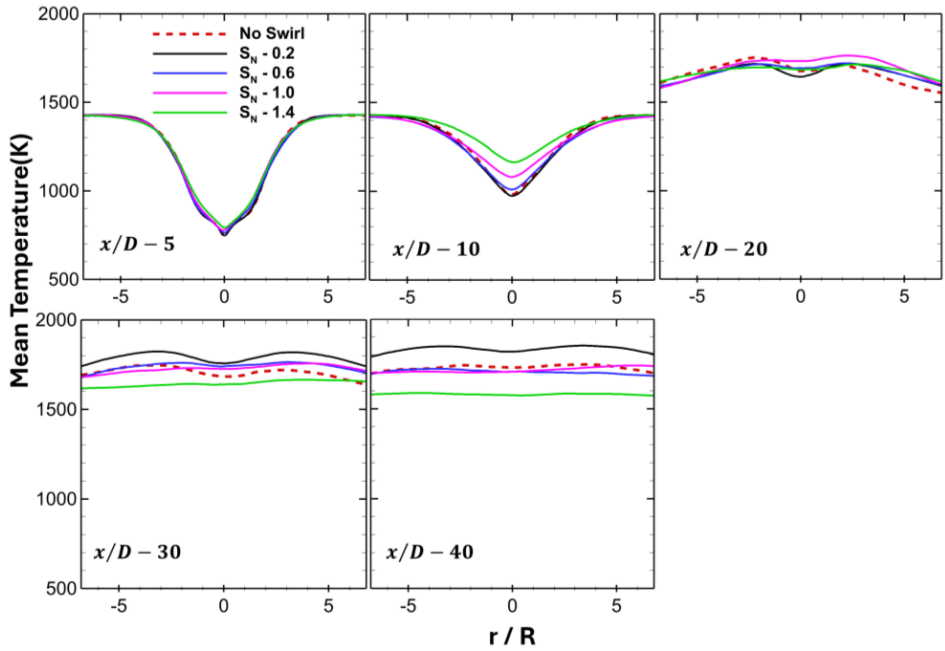

FIG.3 Radial distribution of mean temperature at different downstream locations is compared for four Swirl numbers, i.e. 0.6,1.0,1.0 & 1.4, with no-swirl flames.

Figure 4 displays the mean temperature contours for S1, S2, S3, and S4 flames with Swirl Numbers ($S_N$) 0.2, 0.6, 1.0 and 1.4, respectively. A jet of cold air and fuel enters the center of the flame cone, indicating that all flames have an "annular" flame structure. The coflow's swirling motion influences the position and shape of the flame macrostructure. The downstream region's flame width increases as the swirl increases due to the improved mixing. An intriguing phenomenon of flame dispersion is witnessed in the higher swirl flames - S2, S3, and S4: the sudden transition from a distinct tubular-flame structure to a more uniform high-temperature zone (marked in Fig. 4). As the swirl number increases, the flame's dispersion increases. For example, in the case of S2, the flame begins to disperse at x/D 45, and as the swirl number increases, the flame disperses earlier, as seen in S3 and S4 flames, where it occurs at x/D 35 and x/D 28, respectively. This phenomenon results in a compact and smaller flame with a higher swirl (S4 being the smallest flame), because of which lower mean temperatures were observed at the downstream distances 30D and 40D for the high swirl cases of S2, S3, and S4 than S1. The increased flow tangential velocities lead to stronger centrifugal forces for the flames with larger swirl numbers. These strong centrifugal forces may lead to downstream flame dispersion. The reduction in the flame size with increasing swirl numbers may also be explained by faster combustion of the evaporated fuel in hot coflow or the early flame extinction resulting in incomplete combustion. Further analysis is carried out in the following sections to understand the underlying causes.

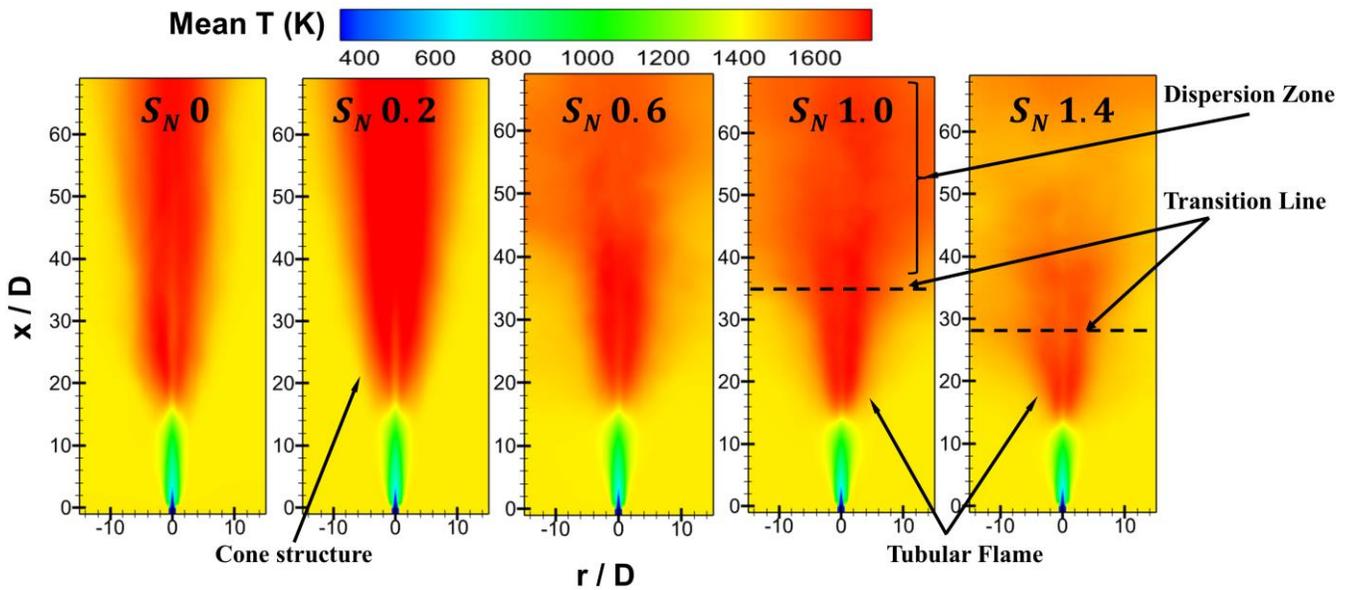
FIG.4 Temperature distribution for 5 flames.

## C. FLAME PROFILE

The hydroxyl (OH) radical is a widely used indicator for flame front analysis, ignition, and local extinction studies. Figure 5 shows the OH contours of the swirl flames, in contrast to a flame with no swirl or $S_N = 0$ (NS-flame). It can be seen that the flame length decreases as the swirl number rises with S2, S3, and S4 flames. Due to higher velocity gradients in the downstream region in high swirl cases, shear layer width increases, improving mixing. As previously mentioned, the S2, S3, and S4 flames exhibit an exciting transition from a narrow, sharper cone to a larger, blunt flame cone. The base of the flame cone indicates the transition from a tubular flame shape to the downstream dispersed combustion zone, as mentioned above. Early flame dispersion is observed in S4-flame (Fig. 4) results from the combustion over a larger region nearer to the burner (with transition point at x/D=28), as indicated by the bigger flame base and larger flame spread.

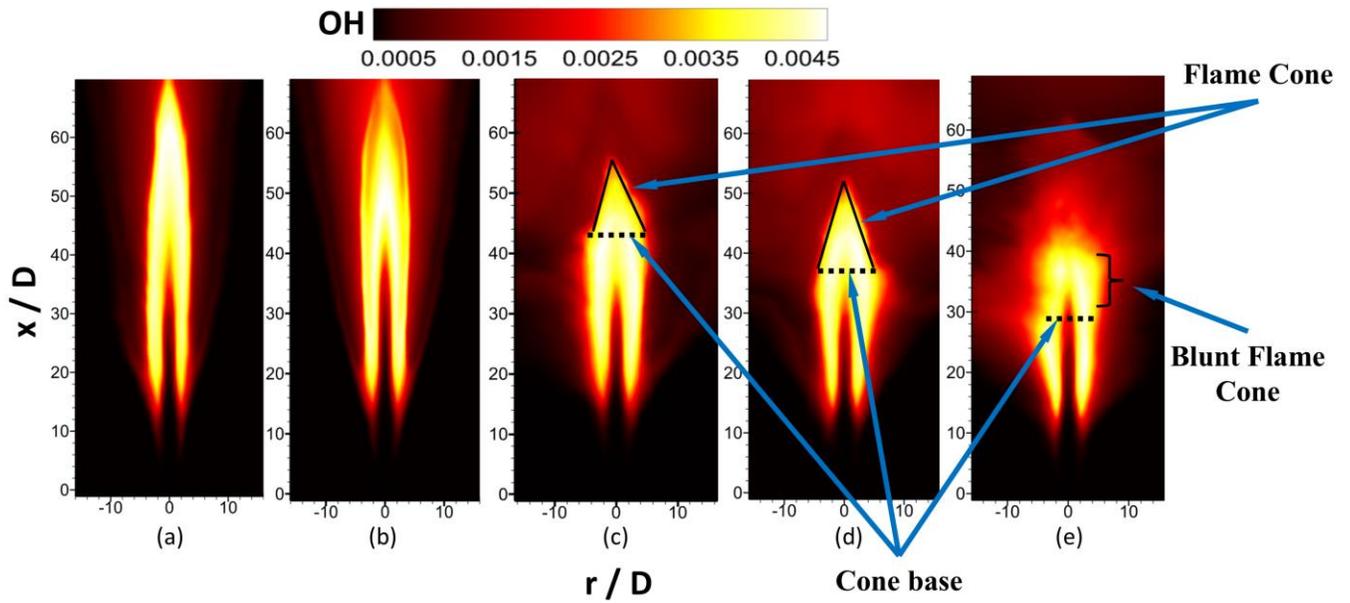

FIG 5. OH radical distribution of 5 flames: (a)$S_N = 0$, (b)$S_N = 0.2$, (c)$S_N = 0.6$, (d)$S_N = 1.0$, and (e)$S_N = 1.4$.

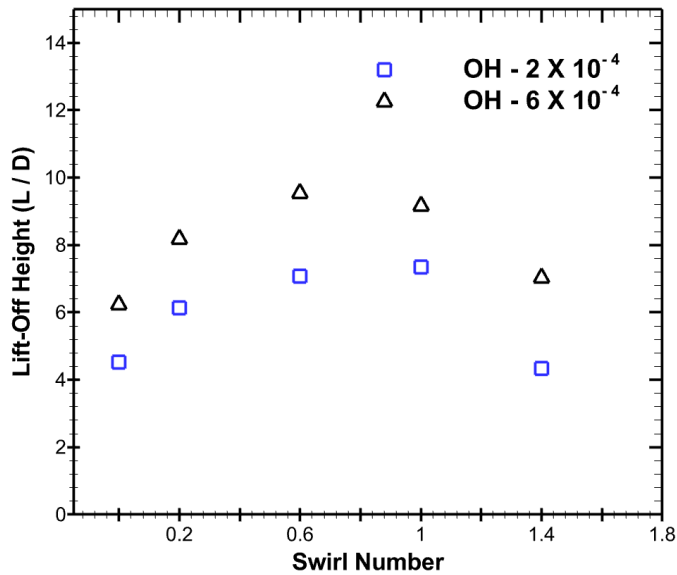

FIG 6. Lift-off height for minimum OH mass fraction thresholds of $2 \times 10^{-4}$ and $6 \times 10^{-4}$.

Another important aspect is the lift-off height and the auto-ignition. Variation in lift-off height can be studied by monitoring the OH concentration as a flame marker. Following the past literature[45] utilizing varying threshold values of the OH mass fraction, this study employs threshold values of $2 \times 10^{-4}$ and $6 \times 10^{-4}$. Although the experimental lift-off height is visually determined from still photos[46], validation for computer analysis necessitates setting particular OH mass fraction values. Two data sets with Favre-averaged OH mass fractions at the designated thresholds are considered for comparison.

The relationship between lift-off height and swirl number is depicted in Figure 6. As the swirl number increases, the lift-off height first rises until S2, after which it falls. The high auto-igniting temperatures near a stoichiometric mixture with meager strain rates are ideal for auto-ignition. Adding a swirl gives the flow tangential momentum without substantially changing the axial flow structure. The overall increased momentum adds to the fluid strain rate and, thus, delays the ignition near the burner. Higher strain rates can potentially elevate the flame farther from the burner and stretch the flame front, increasing its vulnerability to local extinction.

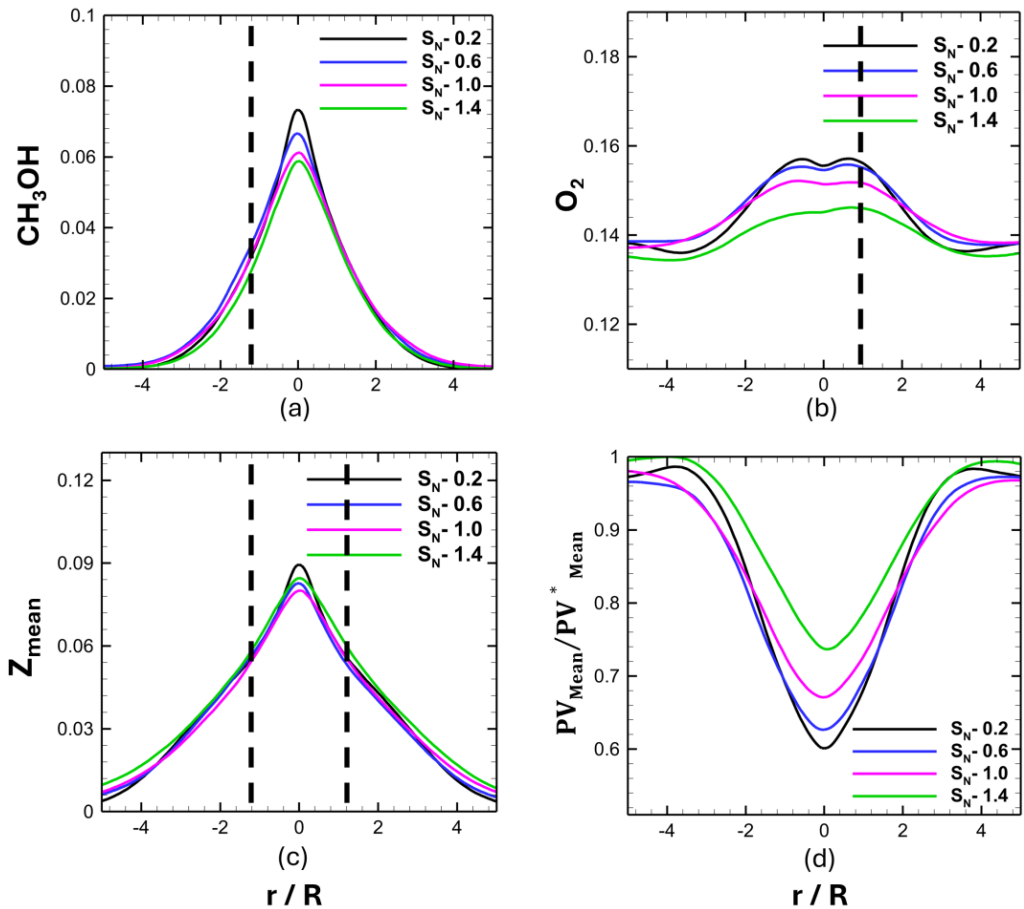

FIG 7. Radial distribution at x/D=8D downstream distance of (a) fuel ($CH_3OH$), (b) oxidizer ($O_2$), (c) Mean Mixture Fraction ($Z_{Mean}$), and (d) Mean progress variable for S2 S3 and S4 flames.

However, the lift-off height drops as the swirl is increased further than $S_N = 0.6$ (S2). The possible reason behind this counter-intuitive drop can be the early development of a near-stoichiometric mixture in the shear region. As visible from Figure 6(a), the presence of liquid droplets in the center jet is why the fuel mass fraction is highest along the jet centerline and drops radially outwards. However, the oxidizer mass fraction is relatively higher in the center jet than in the coflow, though being highest in the shear layer region, as shown in Figure 7(b). The addition of varying degrees of swirl affects the mixing and, consequently, the radial distribution of these reactants. As evident from Figures 7(a) and 7(b), the fuel and oxidizer mass fractions decrease for both the S3 and S4 flames when compared to the corresponding mass fractions of the S2 flame. However, the difference in fuel mass fraction is very small in the shear region of all the flames compared to the oxidizer mass fraction. On comparing the difference in the oxidizer mass fraction of the S3 case from S2 ($\Delta O_{2\to 3}$) and the difference between S3 case from S2 ($\Delta O_{2\to 4}$), it is noted that $\Delta O_{2\to 4} > \Delta O_{2\to 3}$. However, because the mass of the oxidizer decreases with an increasing swirl, the mixture formation approaches the stoichiometry in the shear layer region. Hence, we observe the highest mixture fraction values for S4 flame in the shear layer region, as visible in Figure 7(c), which is an important cause for the decrease of lift-off height. To further consolidate it, the effect of swirl on the droplet statistics will be discussed.

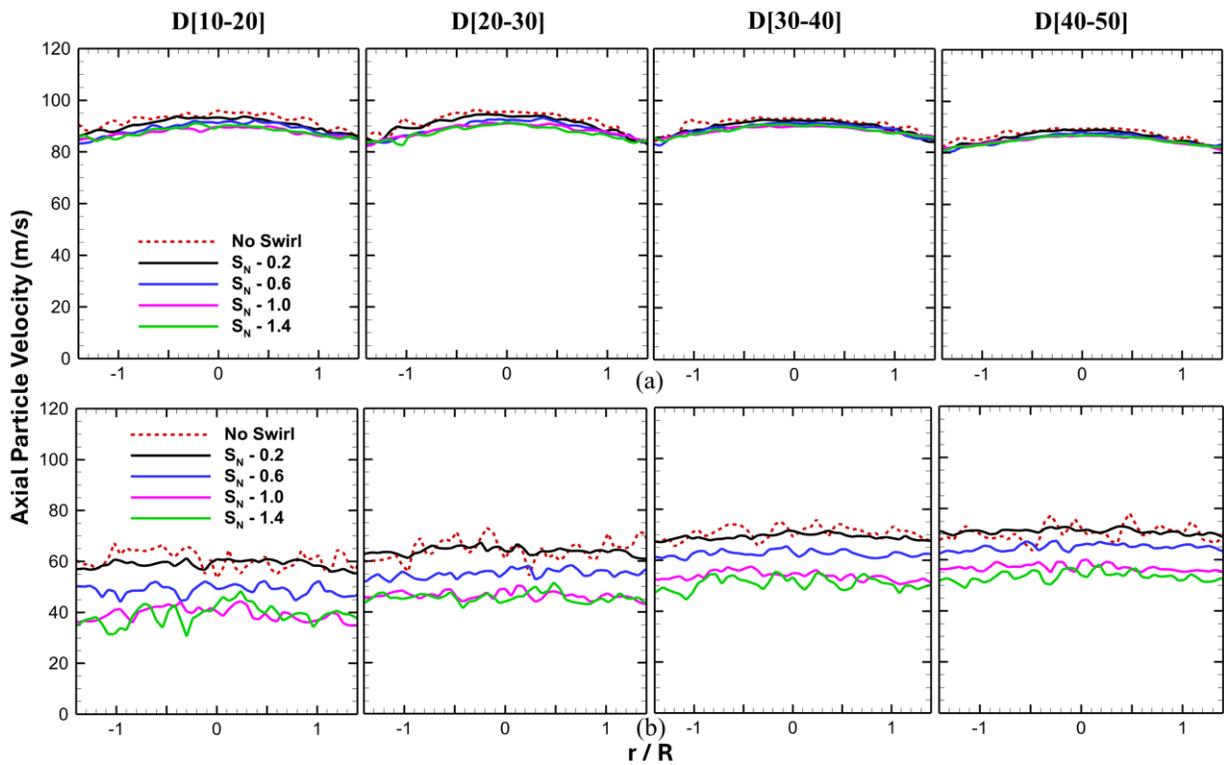

FIG 8. Radial distribution of axial particle velocity at downstream location (a) x/D = 10 and (b) x/D = 30 of $S_N$ = 0.2, 0.6, 1.0 and 1.4 flames compared to no swirl flame.

### D. PARTICLE VELOCITY DISTRIBUTION

The droplet velocity profiles for S1 and S2 flames show a parabolic shape at x/D = 10, as shown in Figure 8(a). Shear forces between the swirling hot coflow and the methanol spray cause the velocity to decrease towards the margins and reach its maximum at the centerline. At this point, there is little size-based distinction, and the velocities of larger and smaller droplets remain comparable. The velocity distribution is mainly dominated by the spray's initial axial momentum at this downstream point, where the swirl's influence is minimal.

The effects are more noticeable with larger swirl numbers (S3, S4). Stronger turbulent mixing causes the droplet velocity profiles to become more flattened over the radial direction at x/D = 10, although they still have a parabolic form. As the increasing swirl improves size-based separation, the velocity differential between larger and smaller droplets becomes more apparent, with smaller droplets more readily influenced by the helical vortices formed due to swirling coflow.

Figure 8(b) shows notable distinctions between these flames that appear by x/D = 30. Swirl flow in the S2 case tends to entrain the droplets and change their velocity by reducing axial components and imparting tangential momentum to them (in the form of the droplet drag forces). S3 and S4 are examples of even higher swirl numbers that intensify these effects even more. Strong centrifugal and turbulence forces increase velocity decay at these swirl levels, making the swirl effect more noticeable. Again, the small droplets are the most affected because of their lower inertia.

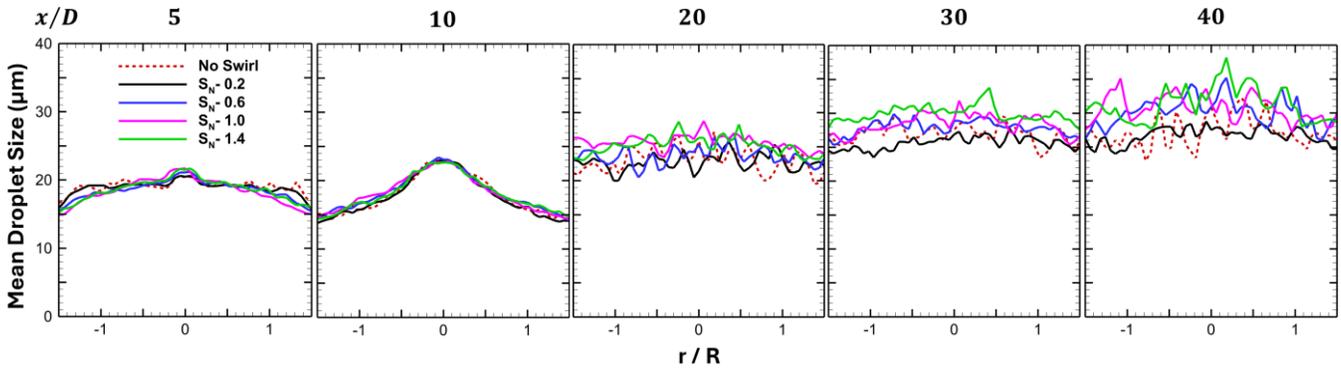

FIG 9. Radial distribution of mean droplet size at different downstream locations of $S_N$ 0.2, 0.6, 1.0 and 1.4 flames compared to no swirl flame.

### E. PARTICLE SIZE DISTRIBUTION

Figure 9 illustrates how the evaporation dynamics driven by the high-temperature coflow surrounding the cold core jet cause the droplet size distribution to vary as the spray flows downstream from the jet exit (from x/D = 5 to x/D = 40). A smoother and more consistent droplet size distribution is produced near the jet exit, at points like x/D = 5 to x/D = 10, where the droplets have less exposure to the hot surroundings and maintain their initial sizes. On the other hand, smaller droplets evaporate more quickly when exposed to high-temperature gases due to their greater surface area-to-volume ratios as the distance increases to x/D = 20 and beyond. The proportion of larger droplets downstream increases relative to the number of small droplets due to the small droplet's easier entrainment, leading to higher evaporation. The larger droplets can stay farther along the flow path because of their higher thermal/mass inertia and slower evaporation rate.

As a result of the reduced presence of tiny droplets and the restricted availability of larger droplets for sampling, the droplet size distribution becomes rougher at further downstream distances. As the spray moves downstream, this causes a less even distribution and an apparent increase in average droplet size. The same reason causes the average droplet size to increase with a higher swirl at a particular downstream distance. As the swirl number increases beyond S2, centrifugal forces expand the gas outwards. The particles with more inertia than the gas continue to maintain their trajectory in the central jet region, so fuel evaporation is mainly concentrated in the central potential core region (see Fig. 6(a)).

**F. LIFTOFF HEIGHT AND FLAME BASE**

Capturing dynamic phenomena like ignition kernel formation, flame propagation, and liftoff height requires a reduced order analysis of flow fields like OH mass fraction. POD breaks down the flow field into energy-ranked modes and helps to identify the larger flow-controlling vortices from the turbulent structures. The most energetic modes are linked to the dominant vortical structures that aid flame lift-off at low to moderate swirl numbers (S1 to S2). After S2, new dominant modes appear, and lift-off height decreases due to large-scale coherent structures stabilizing the flame. POD helps us to thoroughly understand flame propagation and ignition kernel dynamics under various swirl cases described in the next section.

*1. Proper Orthogonal Decomposition setup*

Two swirl numbers, S1 ($S_N$=0.2) and S3 ($S_N$=1.0), are selected to analyze their effect on the flow field and lift-off height. With a sample frequency of 34469.6 $s^{-1}$, we ensure that the small, turbulent structures and larger vortices are captured. As previously discussed, POD is applied to the OH, velocity, and temperature fields for both cases, and three sets of initial data, namely set A (150), B (200), and C (300), are chosen for each field based on the number of samples. In this study, we have chosen set C because it can capture finer flow patterns and provide more precision in predicting the lift-off height behavior. Compared to the S1 flame, the trace of the covariance matrix of OH fluctuations, as shown in equation 22 for the S3 flame, illustrates the more complex igniting dynamics and the enhanced turbulent mixing[38-39].

$$tr(A) = \langle (OH', OH') \rangle = \sum_{k=1}^{N} \lambda_k \qquad (23)$$

*2. Proper Orthogonal Decomposition results*

The first six modes capture around 60% of the total modal energy in the no-swirl (NS-flame) case[24], whereas the first 10 modes account for approximately 45% of the total modal energy in the S1 flame. On the other hand, roughly 65% of the total modal energy for the S3 flame was captured by the first ten modes, as shown in Figure 10(a)(b).

The flow field is dominated by coherent, large-scale structures with few turbulent fluctuations, as evidenced by the high percentage of variance captured by the first six modes in NS-flame. A low swirl in the S1 flame increases the angular momentum, intensifies turbulence, and produces smaller-scale eddies. Energy is scattered across a wide range of these scales, resulting in a dispersed energy distribution. The S1 flame lacks prominent coherent structures, which causes energy to be distributed over more modes. In contrast, the NS and S3 flames feature dominant large-scale structures, increasing variation collected by fewer modes.

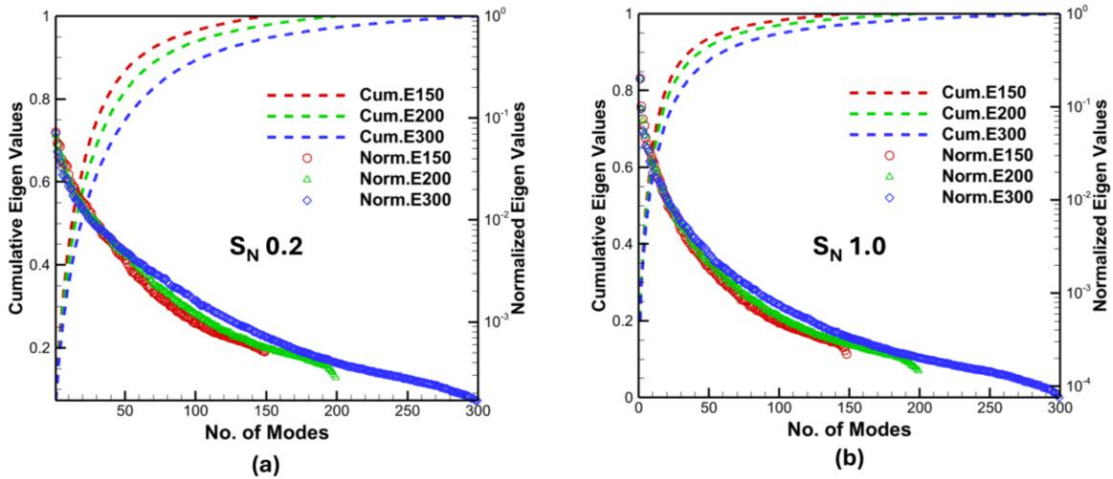

FIG 10. Energy of the eigenmodes from POD analysis of OH field fluctuations for the (a) $S_N = 0.2$ (S1) flame and (b) $S_N = 1.0$ (S3) flame.

An FFT (Fast Fourier Transform) analysis of the fluctuating OH fields for the first ten modes is carried out for the S1 and S3 flames in order to better understand the dominant flow structures and how their periodic behavior causes the liftoff height to first increase and decrease. The first mode of the FFT for the S1 flame, as seen in Figure 11(a), shows a dominant base frequency peak at 67.3 Hz, indicating that this frequency is the main structure in the flow field and is probably connected to large-scale vortical structures. Peaks in the second and third modes are located at 403.9 Hz, suggesting higher-frequency oscillations that are probably connected to coherent flow characteristics and secondary turbulent structures. A systematic interaction between larger and smaller reacting structures is noticeable in this frequency pattern, showing a harmonic link between the first mode and higher modes.

The peaks in the fourth and fifth modes indicate other small-scale structures in the flow, which appear at 134.6 Hz and 471.2 Hz, respectively. The sixth and seventh modes' peaks at 538.6 Hz show finer structures, denoting more significant high-frequency oscillations. The frequency peaks of the ninth and tenth modes are at 269.3 Hz, while the peak of mode 8 is at 336.6 Hz. The frequency distribution shows that the flow is dominated by periodic OH structures, with finer structures being indicated by higher-frequency modes. It should be noted again that all the frequencies indicated above are the higher harmonics of the base frequency of 67.3Hz. Indicative of coherent flame structures, the observed frequencies imply that the S1 flame shows a complicated yet well-organized and intense periodic behavior in the OH radical field.

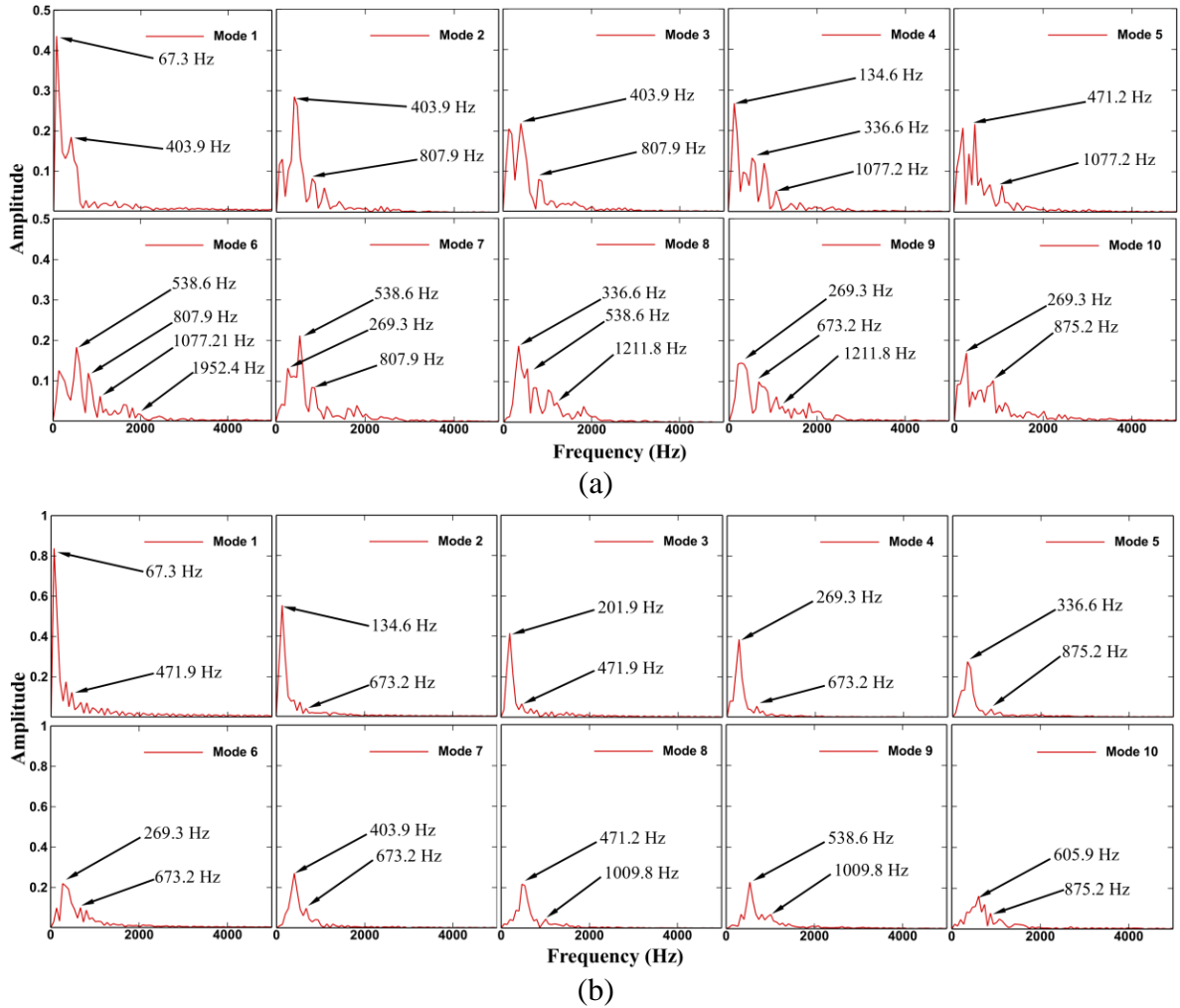

FIG 11. FFT of the time coefficient of first 10 modes of OH field (a) $S_N = 0.2$ flame (b) $S_N = 1.0$ flame.

Like the S1 flame, the S3 flame's FFT analysis in Figure 11(b) likewise displays a dominating base frequency peak for the first mode at 67.3 Hz, indicating that the primary structure is constant for both swirl numbers. The frequency maxima of the second and third modes, which appear at 134.6 Hz and 201.9 Hz, respectively, indicate the existence of secondary structures brought on by more intense swirl.

Peaks at 269.3 Hz and 336.6 Hz are seen in the fourth and fifth modes, suggesting that more important flow structures are interacting and causing more intricate combustion dynamics. The wider variety of scales involved in the combustion process at greater swirl numbers is further demonstrated by the sixth and seventh modes, which show peaks at 269.3 Hz and 403.9 Hz, respectively. The ninth and tenth modes exhibit 538.6 Hz and 605.9 Hz, respectively, while mode 8 peaks at 471.2 Hz.

It should also be noted that the turbulent structures, usually having frequencies more than $5 \times 10^3$, are not visible in the OH analysis. An increased molecular viscosity of the hot gases will lead to turbulence attenuation in the region of the flame. Therefore, the flame will exist in/around the vortical structures of varying frequencies. Despite the increased swirl, energy in the S3 flame is concentrated in dominating low-frequency modes associated with these coherent structures. When the swirl grows from S1 to S3, these coherent structures stabilize the flow, as seen by the shift in the FFT analysis from higher to lower frequency peaks. It is interesting to see what role turbulence plays in the non-reacting zones for the two cases of swirl flames. This illustrates the exciting finding that the higher swirl numbers lower liftoff height by fostering flame stability by forming ordered flow patterns.

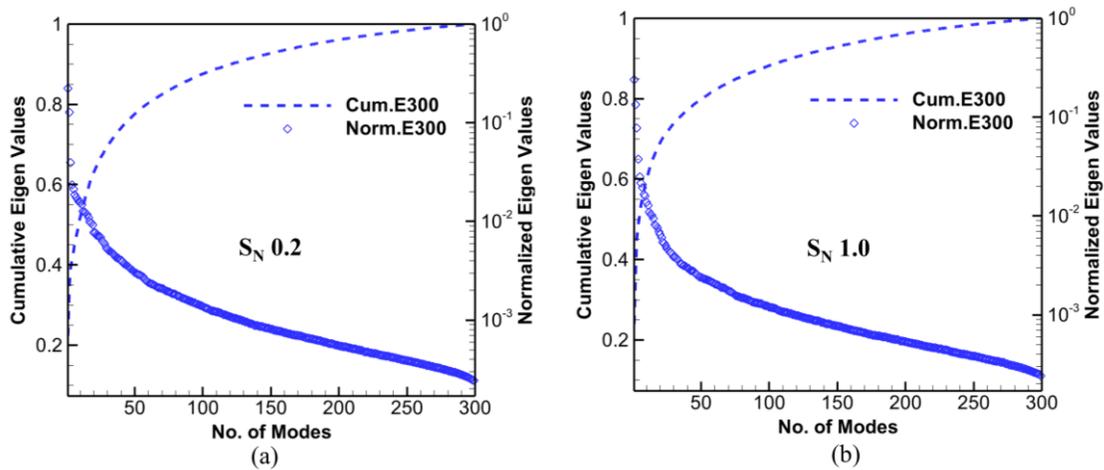

FIG 12. Energy of the eigenmodes from velocity-temperature based POD of (a) $S_N = 0.2$ flame & (b) $S_N = 1.0$ flame.

The instantaneous OH fields and the 10th POD reconstructed OH fields become more chaotic due to greater turbulence and mixing caused by stronger centrifugal forces. Because of the higher turbulence, the S3 flame generates its ignition kernel earlier than the S1 flame. Despite these changes in ignition time and flame structure, the harmonic frequency of the dominant modes remains constant at 67.3 Hz in all cases, indicating that the fundamental periodic nature of the flow remains unchanged.

To answer the effect of flowfield in the non-reacting zones and to assess the effect of vortical structures on flame propagation in high-temperature shear flows, POD of velocity-temperature based flow field is performed for S1 and S3 flames. Regarding the POD of the velocity-temperature field, the covariance matrix indicates the fluctuation energy of vortices dominating the flow. As illustrated in Figure 12, the first 10 modes in the S1 flame capture roughly 52% of the total fluctuation energy, whereas in the S3 flame with a larger swirl number, they capture almost 60%. Stronger, more coherent vortices are formed due to the enhanced swirl in the S3 flame, which amplifies the swirling motion and increases flow instabilities.

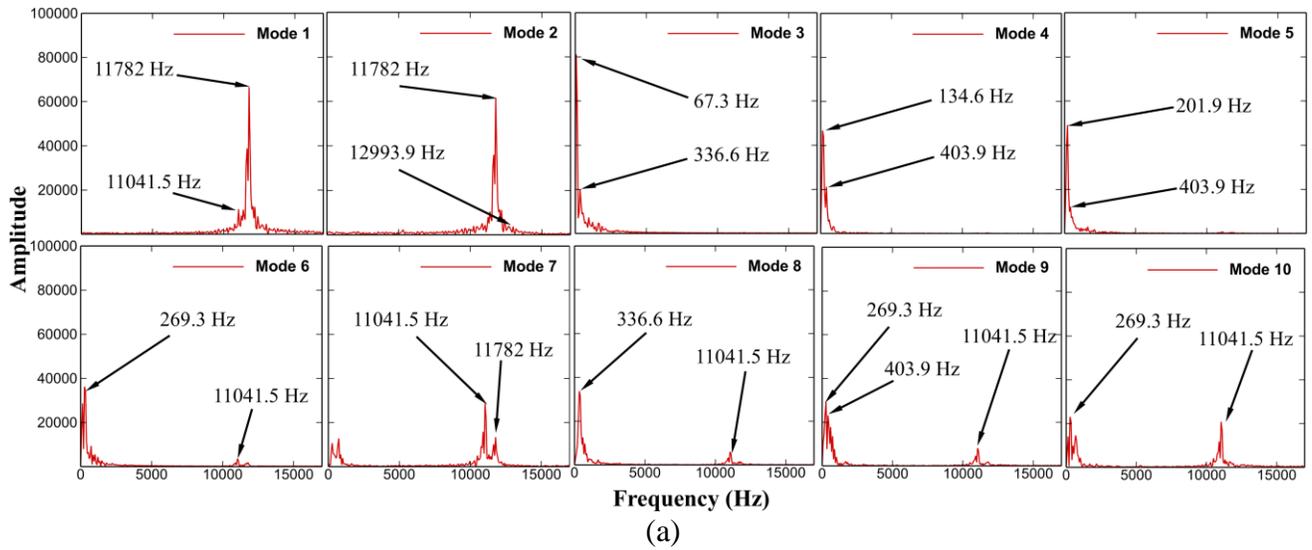

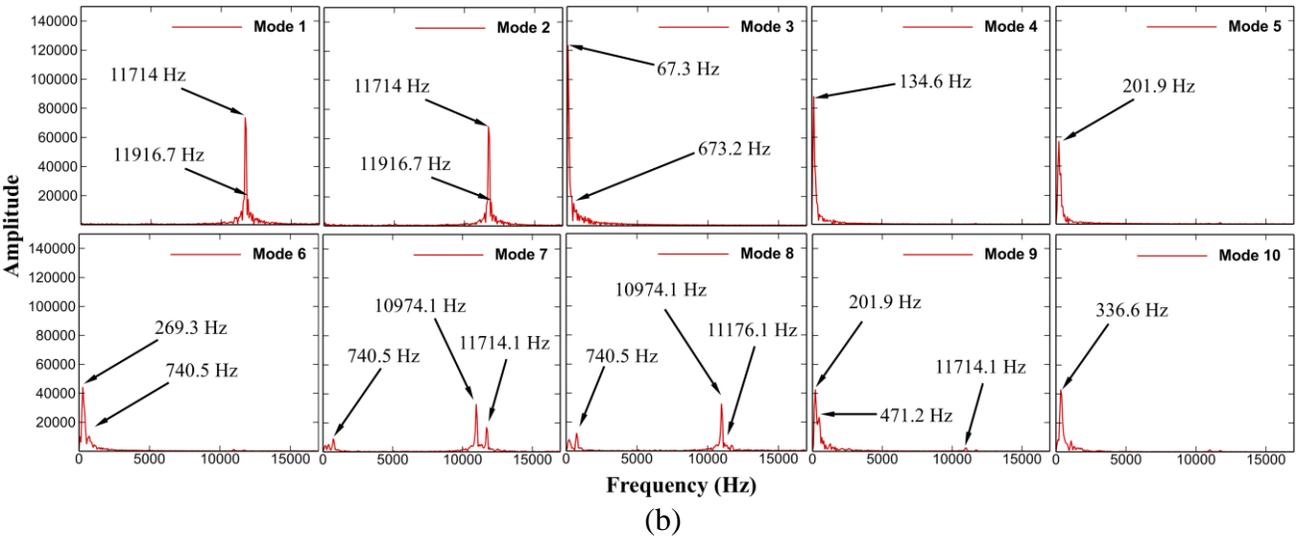

FIG 13. FFT of the time coefficient of first 10 modes from velocity-temperature based POD for (a) $S_N = 0.2$ flame & (b) $S_N = 1.0$ flame.

Different frequency patterns that correlate to various turbulent structures across the first ten modes are revealed by the FFT analysis for S1 and S3 flames, as seen in Figure 13(a) and (b). High-frequency peaks at approximately 11700 Hz are visible in modes 1, 2, 7, and 8 of both the swirl cases. This indicates the dominating presence of turbulence in the non-reacting zones and the breakdown of larger eddies downstream. The low-frequency peak of the third mode at 67.3 Hz is consistent with fluctuating OH field-based POD findings and represents larger, coherent vortices in the upstream or midstream. The other frequencies (134.6 Hz, 201.9 Hz, and 269.3 Hz) in modes 4 to 10 show vortex interactions and periodic nature, typical of medium-scale structures.

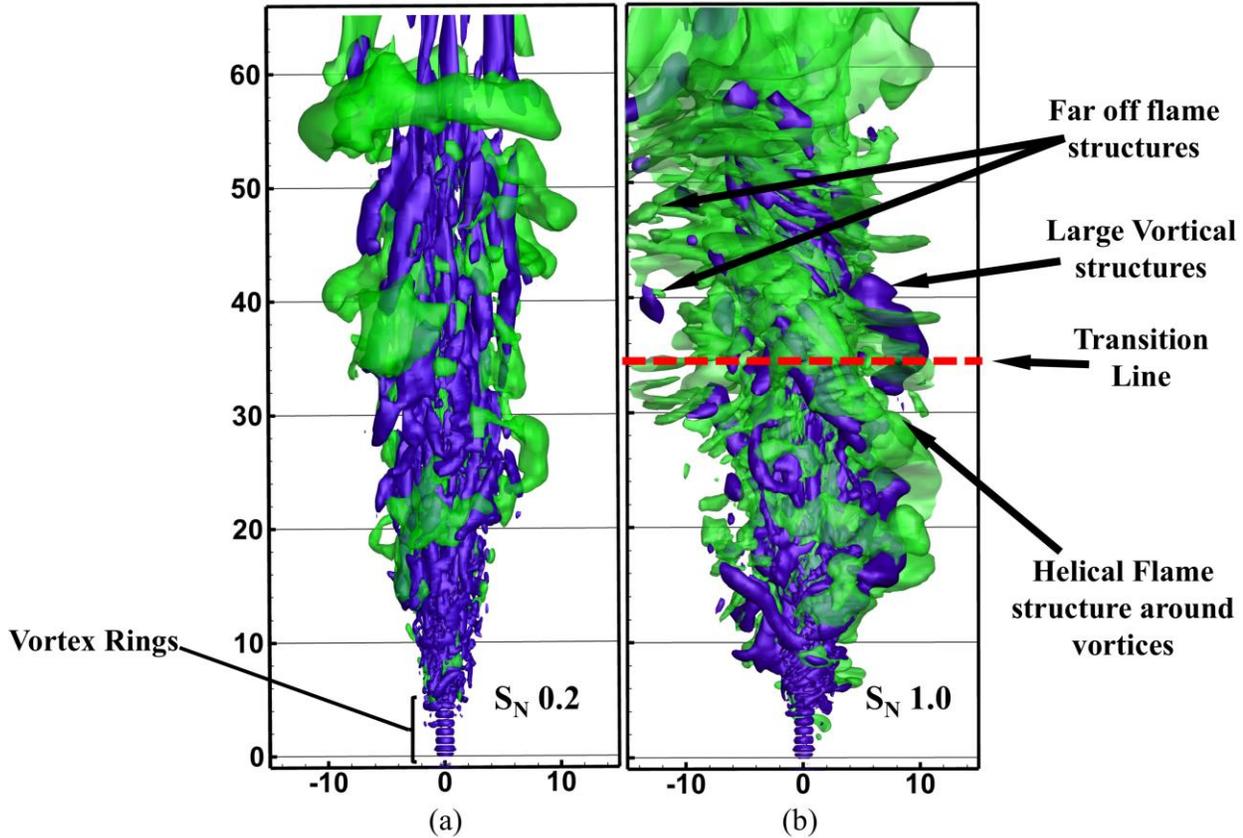

FIG 14. The reconstructed third dominant mode of velocity-temperature POD shows ten-mode flame propagation around vortex structures for (a)$S_N = 0.2$ and (b)$S_N = 1.0$ for the 200$^{th}$ snapshot. Q-criterion visualizes vortex structures at $1\times10^5$, with POD-reconstructed OH fluctuations at $2\times10^{-5}$.

The ten modes of OH mass fraction are used to reconstruct the flame, while a single 3$^{rd}$ mode of velocity-temperature field is used for Q-criterion field reconstruction at the 200$^{th}$ snapshot that provides detailed information of the corresponding structures in a physical space (see figure 14). About 52% of the fluctuation energy is accounted for by the 10 main POD modes for the S1 flame, as shown in Figure 14(a). Ignition kernels form 5D to 10D downstream due to vortical structures, which are less noticeable. Although they quickly evaporate, these larger streamwise vortices resemble broken rings and are crucial for flame propagation near the shear layer due to their intermediate swirl strength. However, because of the S3 flame's higher swirl number, the first 10 POD modes possess around 60% of the fluctuation energy, indicating lesser turbulent structures. However, the stronger swirl results in dominant PVCs (the helical vortices) with a surrounding helical flame (both having a frequency of 67.3 Hz), as shown in Figure 14(b). It is interesting to note that a high swirl case S3 stabilizes the flame with a strong helical vortex while improving the mixing of coflow with the central jet. This is apparent in the averaged plots of velocity, temperature, and mean mixture fraction (Figures 2,3 and 7).

As discussed earlier in Section IV(B), a transition line (approximately at x/D=35) exists above which the flame dispersion is observed, and the tubular flame undergoes a transition to a more uniform, more expansive combustion zone. Figure 14(b) shows that the flame structures move away from the centerline due to the acting centrifugal forces. These far-off flame structures are why the flame shows dispersion phenomena in the averaged flow field.

Numerous crucial elements of a dilute spray flame in an open, fluctuating swirl hot coflow environment have been investigated, but more factors need to be looked at in future studies. The flame dispersion zone at high swirl intensity is critical to combustion efficiency. The newly emerging methodology, like the embedded DNS analysis[47-48], can potentially be used as a future study to investigate the swirl effects on mixing, turbulence, and flame height in this zone.

## V. CONCLUSION

The present study demonstrates the effect of swirl number on a dilute methanol spray flame's lift-off height, ignition, and flame dynamics. The increase in a swirl from low to moderate levels but less than 0.6 pushes the flame base downstream and raises the lift-off height. Further increase in the swirl intensity reduces the lift-off height. Also, the flames with swirl intensity of 0.6 or more showed flame dispersion in the Favre averaged OH fields downstream from the jet exit. This transition is marked by a change in tubular-shaped flame into a more uniform combustion region covering a larger radial distance. The transition point decreases with an increase in the swirl number. Based on these observations, the radial distribution of mean flow fields is analyzed at various downstream distances and studied using proper orthogonal decomposition. Based on these key findings using the above techniques, we infer the following conclusions:

1) For the moderate increase in swirl values until 0.6, the corresponding rise in fluid strain rate delays the ignition and increases the lift-off height.
2) With a further increase in swirl intensity to values more than 0.6, the mixing starts to play a dominant role. It is found that the mixture fraction rises in the shear layer region, consequently reducing the lift-off height.
3) The intense mixing is noticeable in the mean velocity, temperature, and particle velocity distributions for higher swirl values. It also explains the rise in droplet size measured at a particular downstream distance for increasing swirl intensity.
4) The transition to flame dispersion obtained in the averaged OH field for flames at high swirl is due to the rotating flame structures above a particular downstream distance. When these small-scale flame structures, rotating along the jet centerline with the helical flame structure, are averaged, the mean-field gives a uniform distribution over a wide radial distance (referred to as flame dispersion in the present study)
5) Turbulence dominates the flow evolution of flames with low-to-moderate swirl numbers, while more coherence is observed in vortical (like PVCs) and flame structures at high swirl intensities.

Many essential aspects of a dilute spray flame in an open, varying swirl coflow condition are studied, yet other parameters remain to be further examined. Also, the proper orthogonal decomposition has proven to be helpful in explaining the critical aspects of flame-vortices interaction by decomposing the complex turbulent swirl flame into lower-order modes.

**DATA AVAILABILITY**

The data that support the findings of this study are available from the corresponding author upon reasonable request.

**ACKNOWLEDGMENTS**

We acknowledge the National Supercomputing Mission (NSM) for providing computing resources of 'PARAM Sanganak' at IIT Kanpur, which is implemented by C-DAC and supported by the Ministry of Electronics and Information Technology (MeitY) and Department of Science and Technology (DST), Government of India. Also, we would like to thank the computer center ( www.iitk.ac.in/cc ) at IIT Kanpur for providing the resources to carry out this work.

**APPENDIX A: LAGRANGIAN FRAMEWORK**

In equation 8, the effect of turbulence on the particles is modeled using the gradient dispersion model, which is represented by the term $\boldsymbol{F}_T$. Given as follows:

$$F_D = C_D \left( \pi D_p^2 / 8 \right) \rho_g (u_g - u_p) |u_g - u_p| \quad (24)$$

$$F_G = m_p g \left( 1 - \frac{\rho_g}{\rho_p} \right) \quad (25)$$

In the equation for the gravitational force, $F_G$ represents the combined effects of buoyancy and gravity, while $u_p$ and $u_g$ denote the velocities of the Lagrangian parcels and the gas phase, respectively. The diameter of the particles is represented by $D_p$. The drag coefficient of the droplets, $C_D$, is calculated using the Schiller-Naumann equation[29] as:

$$C_D = \begin{cases} 24(1 + 0.15 Re_p^{0.687})/Re_p, & Re_p \leq 1000 \\ 0.44, & Re_p > 1000 \end{cases} \quad (26)$$

$$Re_p = \frac{\rho_g |\tilde{u} - u_p| D_p}{\mu_g} \quad (27)$$

where $Re_p$ is droplet Reynolds number based on the slip velocity and and $\mu_g$ is the gas phase dynamic viscosity.

Assuming spherical droplets, the standard Ranz-Marshal[30] and Frossling[35] correlations are employed for convective heat transfer and mass transfer.

$$\text{Nu} = 2 + 0.552 Re_p^{0.5} \text{Pr}^{0.33} \tag{28}$$

$$\text{Sh} = 2 + 0.552 Re_p^{0.5} \text{Sc}^{0.33} \tag{29}$$

In accordance with Abramzon and Sirignano[49], the Sherwood (Sh) and Nusselt numbers (Nu) are substituted with modified values (denoted by *) to account for the blowing effect brought on by droplet evaporation, which thickens the laminar boundary layer and lowers the transfer rate.

$$Sh^* = 2 + \frac{Sh - 2}{F_M} \qquad\qquad Nu^* = 2 + \frac{Nu - 2}{F_T} \tag{30}$$

The corresponding transfer numbers indicated by $F_M$ and $F_T$ are the same universal function.

$$F = (1 + B)^{0.7} \frac{\ln(1+B)}{B} \tag{31}$$

As temperatures rise, the blowing effect becomes more noticeable. The flashing of liquid is also seen as the temperature rises over the boiling point. Therefore, the model put out by Zuo et al.[50] is based on a mix of evaporation and the flashing process.

$$\dot{m}_e = \frac{\pi k d_0}{c_p}\left(\frac{Nu^*}{1 + \dot{m}_f/\dot{m}_e}\right) \ln\left(1 + \left(1 + \frac{\dot{m}_f}{\dot{m}_e}\right)\frac{h_\infty - h_b}{h_{fg}}\right) \tag{32}$$

The evaporation rate, $\dot{m}_e$, is dependent on the latent heat of vaporization ($h_{fg}$), heat conductivity (k), and heat capacity ($c_p$). Additionally, the enthalpy of gas at the droplet surface and in the gas phase is denoted by $h_b$ and $h_\infty$. The thermodynamic characteristics for the two phases at various pressures and temperatures are computed using the National Institute Standards and Technology[51] NSRDS – AICHE (National Standard Reference Data System – American Institute of Chemical Engineers) database.

**APPENDIX B: FGM FRAMEWORK**

1. **Mixture Fraction ($Z_1$)**

In equation 9, $Y_e$ and $M_w$ are the coupling function and molecular weight of carbon, hydrogen, and oxygen atoms[24].

$$Y_e = 2\frac{Y_C}{M_{w,C}} + 0.5\frac{Y_H}{M_{w,H}} - \frac{Y_O}{M_{w,O}}. \tag{33}$$

### 2. Progress Variable ($Y_c$)

In this equation 10, M represents the molar mass used as the weighting factor. The unscaled progress variable is normalized based on its minimum and maximum values. When considering four control variables, the progress variable's final value is influenced by enthalpy loss, mixture fraction, and a second mixture fraction that characterizes the mixing of the two oxidizer streams[51].

$$C = \frac{Y_c - Y_c^u(Z_1, \eta, Z_2)}{Y_c^b(Z_1, \eta, Z_2) - Y_c^u(Z_1, \eta, Z_2)} \tag{34}$$

Here, 'b' and 'u' indicate $Y_c$ values in burnt and unburnt states.

Also, $h_{ad}$ represents fuel-oxidizer mixture adiabatic enthalpy as:

$$h_{ad} = Z_1 h_f + (1 - Z_1) h_{Ox} \tag{35}$$

In these equations, $h_f$ and $h_{Ox}$ represent the adiabatic enthalpies of the fuel and oxidizer, respectively. The second mixture fraction, $Z_2$, indicates that the oxidizer combines two separate oxidizer streams.

### 3. Enthalpy Deficit

In equation 11, the composition of the oxidizer adiabatic enthalpy is given as follows[24]:

$$h_{ad} = Z_1 h_f + (1 - Z_1) Z_2 h_{Ox_1} + (1 - Z_1)(1 - Z_2) h_{Ox_2} \tag{36}$$

Here, $h_{Ox_2}$ and $h_{Ox_1}$ are hot-coflow and air adiabatic enthalpy, respectively.

**REFERENCES**

[1]Syred, N. (2006). A review of oscillation mechanisms and the role of the precessing vortex core (PVC) in swirl combustion systems. *Progress in Energy and Combustion Science*, *32*(2), 93-161. https://doi.org/10.1146/annurev-fluid-010313-141300
[2]Khalil, A. E., & Gupta, A. K. (2011). Distributed swirl combustion for gas turbine application. Applied Energy, 88(12), 4898-4907. https://doi.org/10.1016/j.apenergy.2011.06.051
[3]Weber, R., & Dugué, J. (1992). Combustion accelerated swirling flows in high confinements. Progress in Energy and Combustion Science, 18(4), 349-367. https://doi.org/10.1016/0360-1285(92)90005-L
[4]Leuckel, W. (1967). Swirl intensities, swirl types and energy losses of different swirl generating devices. International Flame Research Foundation.
[5]Anacleto, P. M., Fernandes, E. C., Heitor, M. V., & Shtork, S. I. (2003). Swirl flow structure and flame characteristics in a model lean premixed combustor. Combustion Science and Technology, 175(8), 1369-1388. https://doi.org/10.1080/00102200302354

[6]Chanaud, R. C. (1965). Observations of oscillatory motion in certain swirling flows. Journal of Fluid Mechanics, 21(1), 111-127. https://doi.org/10.1017/S0022112065000083

[7]Karyeyen, S., Feser, J. S., Jahoda, E., & Gupta, A. K. (2020). Development of distributed combustion index from a swirl-assisted burner. Applied energy, 268, 114967. https://doi.org/10.1016/j.apenergy.2020.114967

[8]Cheng, R. K. (2006). Low swirl combustion. The gas turbine handbook, 241-255.

[9]Paschereit, C. O., Gutmark, E., & Weisenstein, W. (1999). Coherent structures in swirling flows and their role in acoustic combustion control. Physics of Fluids, 11(9), 2667-2678. https://doi.org/10.1063/1.870128

[10]Külsheimer, C., & Büchner, H. (2002). Combustion dynamics of turbulent swirling flames. Combustion and flame, 131(1-2), 70-84. https://doi.org/10.1016/S0010-2180(02)00394-2

[11]Huang, Y., & Yang, V. (2009). Dynamics and stability of lean-premixed swirl-stabilized combustion. Progress in energy and combustion science, 35(4), 293-364. https://doi.org/10.1016/j.pecs.2009.01.002

[12]Sankaran, V., & Menon, S. (2002). LES of spray combustion in swirling flows. Journal of Turbulence, 3(1), 011. https://doi.org/10.1088/1468-5248/3/1/011

[13]Lieuwen, T. C., & Yang, V. (Eds.). (2005). Combustion instabilities in gas turbine engines: operational experience, fundamental mechanisms, and modeling. American Institute of Aeronautics and Astronautics.. https://doi.org/10.2514/4.866807

[14]Zhang, K., Jin, Y., Yao, K., Wang, Y., & Lian, W. (2023). Effects of swirling motion on the cavity flow field and combustion performance. Aerospace Science and Technology, 138, 108275. https://doi.org/10.1016/j.ast.2023.108275

[15]Foust, M., Thomsen, D., Stickles, R., Cooper, C., & Dodds, W. (2012). Development of the GE aviation low emissions TAPS combustor for next generation aircraft engines. In 50th AIAA aerospace sciences meeting including the new horizons forum and aerospace exposition (p. 936). https://doi.org/10.2514/6.2012-936

[16]Heath, C. M. (2014). Characterization of swirl-venturi lean direct injection designs for aviation gas turbine combustion. Journal of Propulsion and Power, 30(5), 1334-1356. https://doi.org/10.2514/1.B35077

[17]Zhu, P., Li, Q., Feng, X., Liang, H., Suo, J., & Liu, Z. (2023). Experimental and Simulation Study on the Emissions of a Multi-Point Lean Direct Injection Combustor. Journal of Applied Fluid Mechanics, 16(10), 1938-1950. https://doi.org/10.47176/jafm.16.10.1863

[18]Jones, W. P., Lyra, S., & Navarro-Martinez, S. (2011). Large eddy simulation of a swirl stabilized spray flame. Proceedings of the Combustion Institute, 33(2), 2153-2160. https://doi.org/10.1016/j.proci.2010.07.032

[19]Dinesh, K. R., Kirkpatrick, M. P., & Jenkins, K. W. (2010). Investigation of the influence of swirl on a confined coannular swirl jet. Computers & Fluids, 39(5), 756-767. https://doi.org/10.1016/j.compfluid.2009.12.004

[20]Rodrigues, H. C., Tummers, M. J., van Veen, E. H., & Roekaerts, D. J. (2015). Spray flame structure in conventional and hot-diluted combustion regime. Combustion and Flame, 162(3), 759-773. https://doi.org/10.1016/j.combustflame.2014.07.033

[21]Prasad, V. N., Masri, A. R., Navarro-Martinez, S., & Luo, K. H. (2013). Investigation of auto-ignition in turbulent methanol spray flames using Large Eddy Simulation. Combustion and flame, 160(12), 2941-2954. https://doi.org/10.1016/j.combustflame.2013.07.004

[22]Heye, C., Raman, V., & Masri, A. R. (2015). Influence of spray/combustion interactions on auto-ignition of methanol spray flames. Proceedings of the Combustion Institute, 35(2), 1639-1648. https://doi.org/10.1016/j.proci.2014.06.087

[23]Sharma, E., De, S., & Cleary, M. J. (2021). LES of a lifted methanol spray flame series using the sparse Lagrangian MMC approach. Proceedings of the Combustion Institute, 38(2), 3399-3407. https://doi.org/10.1016/j.proci.2020.07.088

[24]Bhatia, B., De, A., Roekaerts, D., & Masri, A. R. (2022). Numerical analysis of dilute methanol spray flames in vitiated coflow using extended flamelet generated manifold model. Physics of Fluids, 34(7). https://doi.org/10.1063/5.0098705

[25]Germano, M., Piomelli, U., Moin, P., & Cabot, W. H. (1991). A dynamic subgrid-scale eddy viscosity model. Physics of Fluids A: Fluid Dynamics, 3(7), 1760-1765. https://doi.org/10.1063/1.857955

[26]Yoshizawa, A. (1986). Statistical theory for compressible turbulent shear flows, with the application to subgrid modeling. The Physics of fluids, 29(7), 2152-2164. https://doi.org/10.1063/1.865552

[27]Gadalla, M., Kannan, J., Tekgül, B., Karimkashi, S., Kaario, O., & Vuorinen, V. (2020). Large-eddy simulation of ECN spray A: Sensitivity study on modeling assumptions. Energies, 13(13), 3360.

[28]Greenshields, C.J. "OpenFOAM: the open source CFD toolbox." User Guide (2015).

[29]Naumann, Z., & Schiller, L. J. Z. V. D. I. (1935). A drag coefficient correlation. Z. Ver. Deutsch. Ing, 77(318), e323.

[30]Ranz, W. E., & Marshall, W. R. (1952). Evaporation from droplets. Chem. Eng. Prog, 48(3), 141-146.

[31]Aggarwal, S. K., & Peng, F. (1995). A review of droplet dynamics and vaporization modeling for engineering calculations.

[32]Hermanns, R. T. E. (2001). CHEM1D, a one-dimensional laminar flame code. Report, Eindhoven University of Technology.

[33]Lindstedt, R. P., & Meyer, M. P. (2002). A dimensionally reduced reaction mechanism for methanol oxidation. Proceedings of the combustion institute, 29(1), 1395-1402. https://doi.org/10.1016/S1540-7489(02)80171-7

[34]Bilger, R. W. (2011). A mixture fraction framework for the theory and modeling of droplets and sprays. Combustion and Flame, 158(2), 191-202. https://doi.org/10.1016/j.combustflame.2010.08.008

[35]Frössling, N. "The Evaporating of Falling Drops (in German)". Gerlands Beitrage zur Geophysik, 52:170–216. (1938).

[36]Peters, N. "Turbulent combustion." (2001): 2022. https://doi.org/10.1088/0957-0233/12/11/708

[37]Alam, Z., Patel, R., Bhatia, B., & De, A. (2024, June). Investigation of Methanol Spray Combustion in a Swirling Hot Coflow. In Turbo Expo: Power for Land, Sea, and Air (Vol. 87943, p. V03AT04A020). American Society of Mechanical Engineers. https://doi.org/10.1115/GT2024-122858

[38]Lumley, J. L., & Poje, A. (1997). Low-dimensional models for flows with density fluctuations. *Physics of Fluids*, *9*(7), 2023-2031. https://doi.org/10.1063/1.869321

[39]Meyer, K. E., Pedersen, J. M., & Özcan, O. (2007). A turbulent jet in crossflow analysed with proper orthogonal decomposition. *Journal of Fluid Mechanics*, *583*, 199-227. https://doi.org/10.1017/S0022112007006143

[40]Procacci, A., Kamal, M. M., Mendez, M. A., Hochgreb, S., Coussement, A., & Parente, A. (2022). Multi-scale proper orthogonal decomposition analysis of instabilities in swirled and stratified flames. Physics of Fluids, 34(12). https://doi.org/10.1063/5.0127956

[41]Kypraiou, A. M., Dowling, A., Mastorakos, E., & Karimi, N. (2015). Proper orthogonal decomposition analysis of a turbulent swirling self-excited premixed flame. In 53rd AIAA Aerospace Sciences Meeting (p. 0425).

[42]O'Loughlin, W., & Masri, A. R. (2012). The structure of the auto-ignition region of turbulent dilute methanol sprays issuing in a vitiated co-flow. Flow, turbulence and combustion, 89, 13-35.

[43]Syred, N., & Beer, J. M. (1974). Combustion in swirling flows: a review. *Combustion and flame*, *23*(2), 143-201. https://doi.org/10.1016/0010-2180(74)90057-1

[44]Celik, I. B., Cehreli, Z. N., & Yavuz, I. "Index of resolution quality for large eddy simulations." (2005): 949-958. https://doi.org/10.1115/1.1990201

[45]De, S., De, A., Jaiswal, A., & Dash, A. "Stabilization of lifted hydrogen jet diffusion flame in a vitiated co-flow: Effects of jet and coflow velocities, coflow temperature and mixing." International journal of hydrogen energy 41, no. 33 (2016): 15026-15042. https://doi.org/10.1016/j.ijhydene.2016.06.052

[46]Cabra, R., Myhrvold, T., Chen, J. Y., Dibble, R. W., Karpetis, A. N., & Barlow, R. S. ("Simultaneous laser Raman-Rayleigh-LIF measurements and numerical modeling results of a lifted turbulent H2/N2 jet flame in a vitiated coflow." Proceedings of the Combustion Institute 29, no. 2 (2002): 1881-1888. https://doi.org/10.1016/S1540-7489(02)80228-0

[47]Domingo, P., & Vervisch, L. (2023). Recent developments in DNS of turbulent combustion. Proceedings of the Combustion Institute, 39(2), 2055-2076. https://doi.org/10.1016/j.proci.2022.06.030

[48]Gadalla, M., Karimkashi, S., Kabil, I., Kaario, O., Lu, T., & Vuorinen, V. (2024). Embedded direct numerical simulation of ignition kernel evolution and flame initiation in dual-fuel spray assisted combustion. Combustion and Flame, 259, 113172. https://doi.org/10.1016/j.combustflame.2023.113172

[49]Abramzon, B., & Sirignano, W. A. (1989). Droplet vaporization model for spray combustion calculations. International journal of heat and mass transfer, 32(9), 1605-1618. https://doi.org/10.1016/0017-9310(89)90043-4

[50]Zuo, B., Gomes, A. M., & Rutland, C. J. (2000). Modelling superheated fuel sprays and vaproization. International Journal of Engine Research, 1(4), 321-336. https://doi.org/10.1243/1468087001545218

[51]Daubert, T. E. (1989). Physical and thermodynamic properties of pure chemicals: data compilation. Design Institute for Physacal Property Data (DIPPR).